%% file: ExpFinder_SIMPA.tex
\documentclass[preprint]{elsarticle}

\usepackage{amssymb}
\usepackage{hyperref}
\usepackage{url}
\usepackage{amsmath}
\usepackage{amssymb}
\usepackage{amsthm}
\usepackage{mathrsfs}
\usepackage{graphicx}
\usepackage{todonotes}
\usepackage{multirow}
\usepackage{array}
\usepackage[linesnumbered,ruled,vlined]{algorithm2e}
\usepackage{chngpage}
\usepackage{booktabs}
\usepackage{xcolor}
\usepackage[T1]{fontenc}
\usepackage{tabu}
\usepackage{longtable}
\usepackage{xspace}
\usepackage{microtype}
\usepackage{enumitem}
\usepackage{adjustbox}
\usepackage{wasysym}
\usepackage{subcaption}
\usepackage{geometry}
\usepackage{blkarray}
\usepackage{cleveref}
\usepackage[normalem]{ulem}
\usepackage{xcolor}
\usepackage{float}

\usepackage{soul,xcolor}
\newcommand{\mc}{\mathcal}

\SetCommentSty{mycommfont}

\setstcolor{red}

\journal{Software Impacts}

\begin{document}

\begin{frontmatter}



\title{An open-source framework for ExpFinder integrating $N$-gram Vector Space Model and $\mu$CO-HITS}


\author[csse]{Hung Du\corref{cor}}
\ead{hungdu@swin.edu.au}

\author[media]{Yong-Bin Kang\corref{cor}}
\ead{ykang@swin.edu.au}
\cortext[cor]{Corresponding author}

\address[csse] {Department of Computer Science and Software Engineering, Swinburne University of Technology, Australia}
\address[media] {Department of Media and Communication, Swinburne University of Technology, Australia}

\begin{abstract}
Finding experts drives successful collaborations and high-quality product development in academic and research domains. To contribute to the expert finding research community, we have developed ExpFinder which is a novel ensemble model for expert finding by integrating an $N$-gram vector space model ($n$VSM) and a graph-based model ($\mu$CO-HITS). This paper provides descriptions of ExpFinder's architecture, key components, functionalities, and illustrative examples. ExpFinder is an effective and competitive model for expert finding, significantly outperforming a number of expert finding models as presented in \cite{kang2021expfinder}.
\end{abstract}

\begin{keyword}
ExpFinder  \sep Expert finding \sep N-gram Vector Space Model \sep µCO-HITS \sep Expert collaboration graph



\end{keyword}

\end{frontmatter}

%\linenumbers

% \noindent
% \textbf{Main text}\\
% Maximum 3 pages (excluding metadata, tables, figures, references)\\

% \noindent
% The description of your software  shall include:
% \begin{itemize}
% \item A short description of the high-level functionality and purpose of the software for a diverse, non-specialist audience
% \item An Impact overview that illustrates the purpose of the software and its achieved results:
% \begin{itemize}
% \item[-] Indicate in what way, and to what extent, the pursuit of existing research questions is improved. 
% \item[-] Indicate in what way new research questions can be pursued because of the software.
% \item[-] Indicate in what way the software has changed the daily practice of its users.
% \item[-] Indicate how widespread the use of the software is within and outside the intended user group.
% \item[-] Indicate in what way the software is used in commercial settings and/or how it led to the creation of spin-off companies (if so).
% \end{itemize}
% \item Mentions (if applicable) of any ongoing research projects using the software. 
% \item A list of all scholarly publications enabled by the software.
% \end{itemize}
\input{intro.tex}

\input{functionality.tex}

\input{examples}

\input{impact.tex}

% \input{future.tex}

% \section{Acknowledgements}
% \label{}

% Optionally thank people and institutes you need to acknowledge. 

\section{Declaration of Competing Interest}

The authors declare that they have no known competing financial interests or personal relationships that could have appeared to influence the work reported in this paper.

% \section*{References}



%% References:
\bibliographystyle{elsarticle-num} 
\bibliography{hdu}




% \bibitem{hdu}

% \end{thebibliography}

% \section*{Illustrative Examples}
% Optional : you may include one explanatory  video that will appear next to your article, in the right hand side panel. (Please upload any video as a single supplementary file with your article. Only one MP4 formatted, with 50MB maximum size, video is possible per article. Recommended video dimensions are 640 x 480 at a maximum of 30 frames / second. Prior to submission please test and validate your .mp4 file at  \url{http://elsevier-apps.sciverse.com/GadgetVideoPodcastPlayerWeb/verification} . This tool will display your video exactly in the same way as it will appear on ScienceDirect. )

\section{Current code version}
\label{}

Ancillary data table required for subversion of the codebase. Kindly replace examples in right column with the correct information about your current code, and leave the left column as it is.

\begin{table}[!h]
\begin{tabular}{|l|p{6.5cm}|p{6.5cm}|}
\hline
\textbf{Nr.} & \textbf{Code metadata description} & \\
\hline
C1 & Current code version & v1.0 \\
\hline
C2 & Permanent link to code/repository used for this code version & \url{https://github.com/Yongbinkang/ExpFinder} \\
\hline
C3  & Permanent link to Reproducible Capsule & \url{https://doi.org/10.24433/CO.0133456.v1} \\
\hline
C4 & Legal Code License & MIT License (MIT)\\
\hline
C5 & Code versioning system used & git \\
\hline
C6 & Software code languages, tools, and services used & Python \\
\hline
C7 & Compilation requirements, operating environments \& dependencies & Python environment version 3.6 or above, pandas, networkx, NumPy, scikit-learn, nltk, SciPy, Torch, Transformers, SciBERT  \\
\hline
C8 & Link to developer documentation/manual & \url{https://github.com/Yongbinkang/ExpFinder/blob/main/README.md} \\
\hline
C9 & Support email for questions & ykang@swin.edu.au, hungdu@swin.edu.au \\
\hline
\end{tabular}
\caption{Code metadata}
\label{} 
\end{table}

\end{document}

%% file: intro.tex
\section{Introduction}

\begin{figure*}[!t]
    \centering
    \includegraphics[trim=0cm 0cm 0cm 0cm, clip, width=350pt]{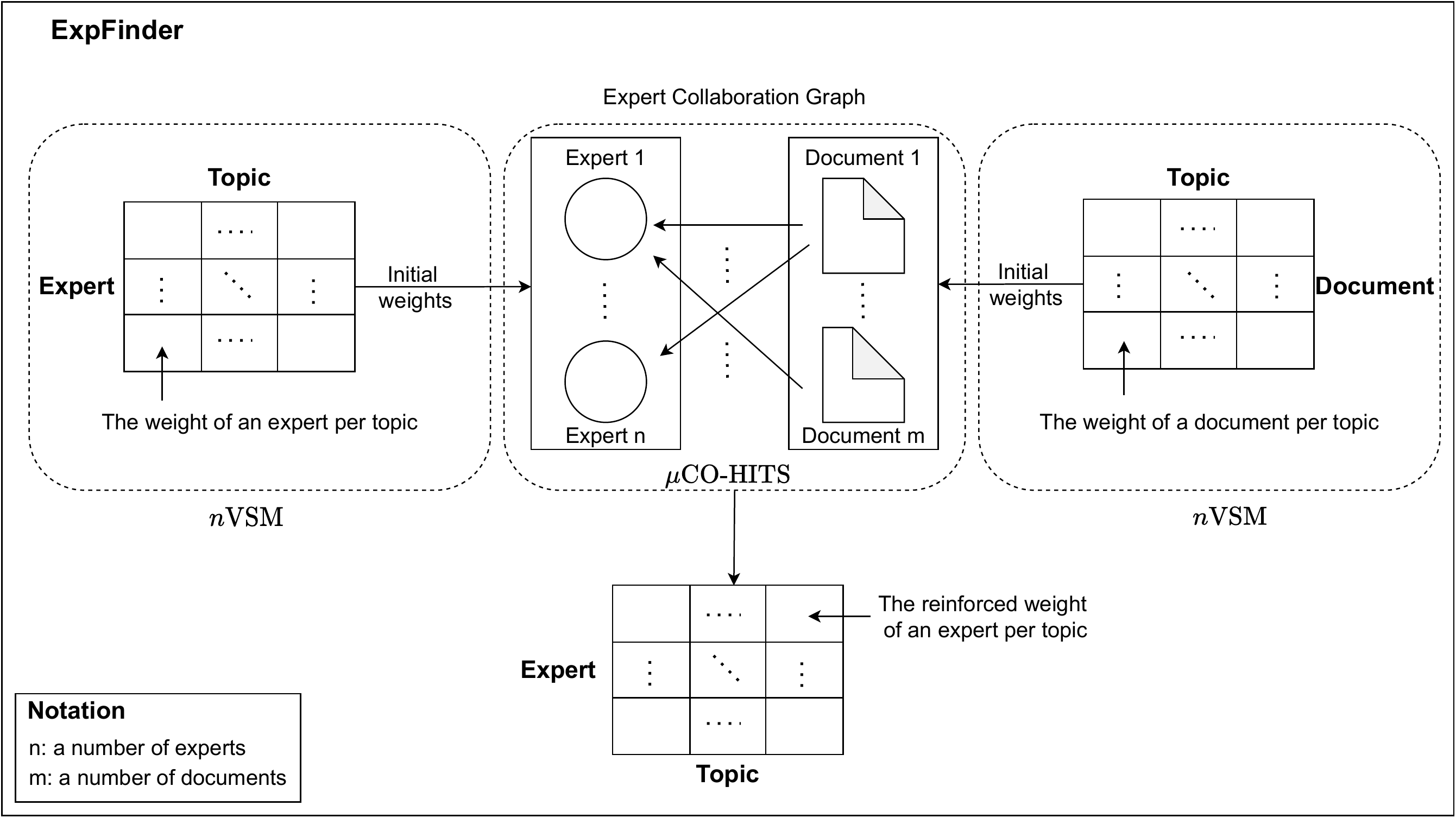}
    \caption{The overview of ExpFinder} 
    \label{fig:intro_method}
\end{figure*}

Identifying experts given a query topic, known as \textit{expert finding}, is a crucial task that accelerates rapid team formation for research innovations or business growth. Existing expert finding models can be classified into three categories such as \textit{vector space models} (VSM) \cite{riahi2012finding, chuang2014combining}, \textit{document language models} (DLM) \cite{Balog2009, Wang2015, WISER2019}, or \textit{graph-based models} (GM) \cite{Hongbo2009, Gollapalli2013, Daniel2015}. ExpFinder \cite{kang2021expfinder} is an ensemble model for expert finding which integrates a novel $N$-gram VSM ($n$VSM) with a GM ($\mu$CO-HITS)-a variant of the generalised CO-HITS algorithm \cite{Hongbo2009}. 

As seen in Figure \ref{fig:intro_method}, ExpFinder has $n$VSM, a vector space model, as a key component that estimates the weight of an expert and a document given a topic by leveraging the Inverse Document Frequency (IDF) weighting~\cite{nidf:2017} for $N$-gram words (simply $N$-grams). Another key component in ExpFinder is $\mu$CO-HITS that is used to reinforce the weights of experts and documents given a topic in $n$VSM using an Expert Collaboration Graph (ECG) that is a certain form of an expert social network. The output of ExpFinder is the reinforced weights of experts given topics. 

ExpFinder is designed and developed to improve the performance for expert finding. In this paper, we highlight two main contributions to the expert finding community. First, we provide a comprehensive implementation detail of all steps taken in ExpFinder. It could also be used as an implementation guideline for developing various DLM-, VSM- and GM-based expert finding approaches. Second, we illustrate how ExpFinder works with a simple example, thus researchers and practitioners can easily understand ExpFinder's design and implementation.

This paper is organised as follows. Section \ref{sec:func} describes ExpFinder's architecture and functionalities. Section \ref{sec:example} demonstrates the procedural steps in ExpFinder. Section \ref{sec:impact} provides the impact and conclusion of ExpFinder.

%% file: functionality.tex
\section{Functionality} \label{sec:func}

ExpFinder is implemented in Python (version $\geq$ 3.6) with open-source libraries such as \texttt{pandas}, \texttt{NumPy}, \texttt{scikit-learn}, \texttt{SciPy},  \texttt{nltk}, and \texttt{networkx}. In this section, we present its architecture, key components, and their functionalities. The architecture of ExpFinder is presented in Figure \ref{fig:func_graph} that consists of four key steps with the corresponding functions and their functional dependencies:

\begin{figure*}[!t]
\centering
\includegraphics[trim=0cm 0cm 0cm 0cm, clip, width=350pt]{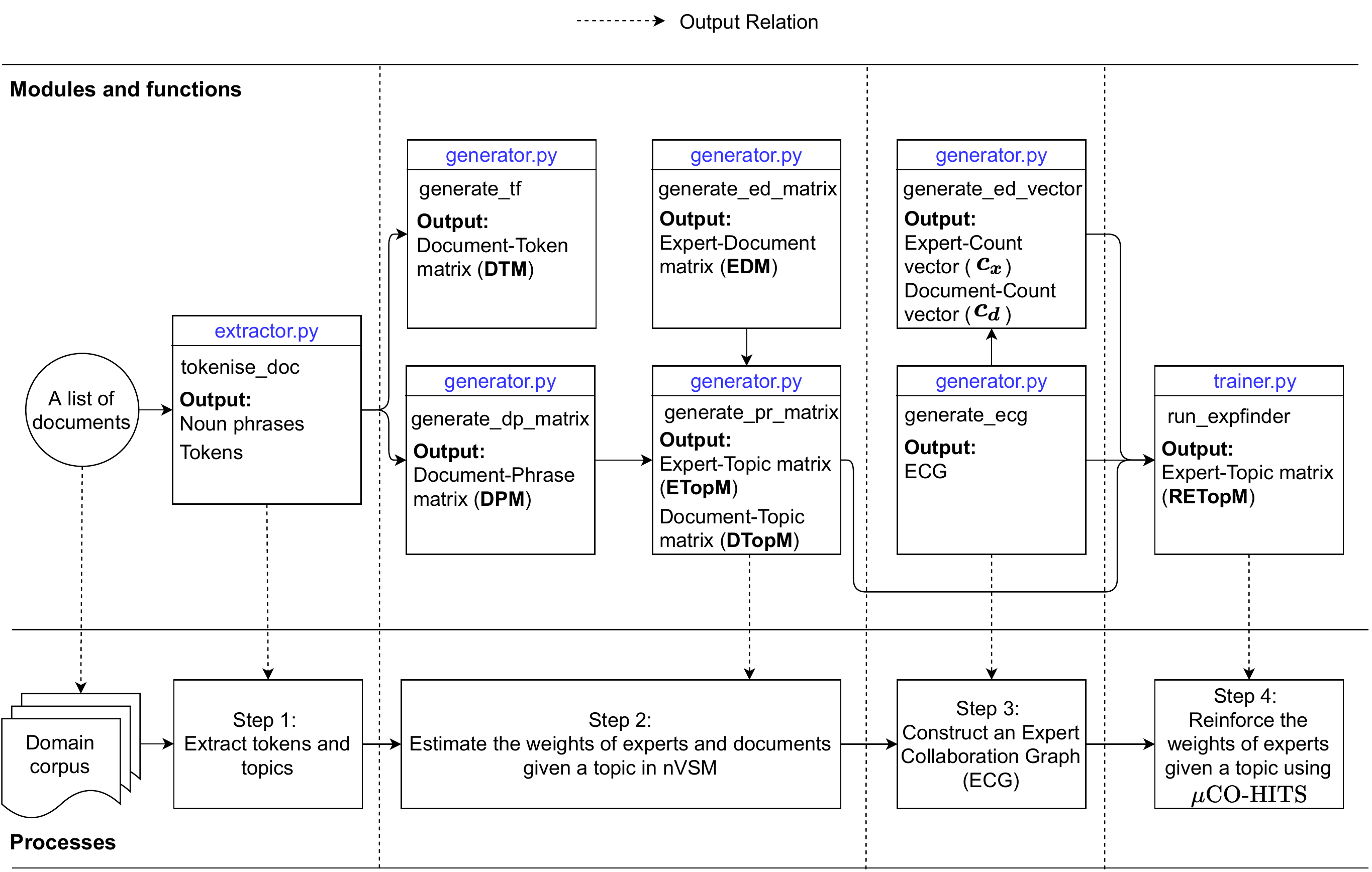}
\caption{The architecture and functional workflow of ExpFinder: blue labels indicate module names of ExpFinder, and `Output Relation' maps the functional component to the corresponding processing step.}
\label{fig:func_graph}
\end{figure*}

\begin{enumerate}
    \item \textbf{Step 1 - Extract tokens and topics}: Given an expertise source $\mc{D}$ (e.g., scientific publications) of experts $\mc{X}$, we extract expertise topics by using  \texttt{tokenise\_doc()} in \textbf{extractor.py}. We assume that expertise topics are represented in the forms of noun phrases. For each document $d\in \mc{D}$, the function splits it into sentences. Then, for each sentence, the function removes stopwords, assigns a part of speech (POS) to each word, merges the inflected forms of a word (i.e., the lemmatisation process, for example, `patients' is lemmatised to `patient'), and extracts single-word terms (called \textit{tokens}) and topics with a given linguistic pattern. In addition, we use a regular expression (\texttt{regex}) in Python to construct a linguistic pattern based on POS that is further used for extracting four different types of topics as shown in Figure \ref{fig:regex}. 
    \begin{figure}[!h]
    \centering
    \includegraphics[trim=0cm 0cm 0cm 0cm, clip, width=250pt]{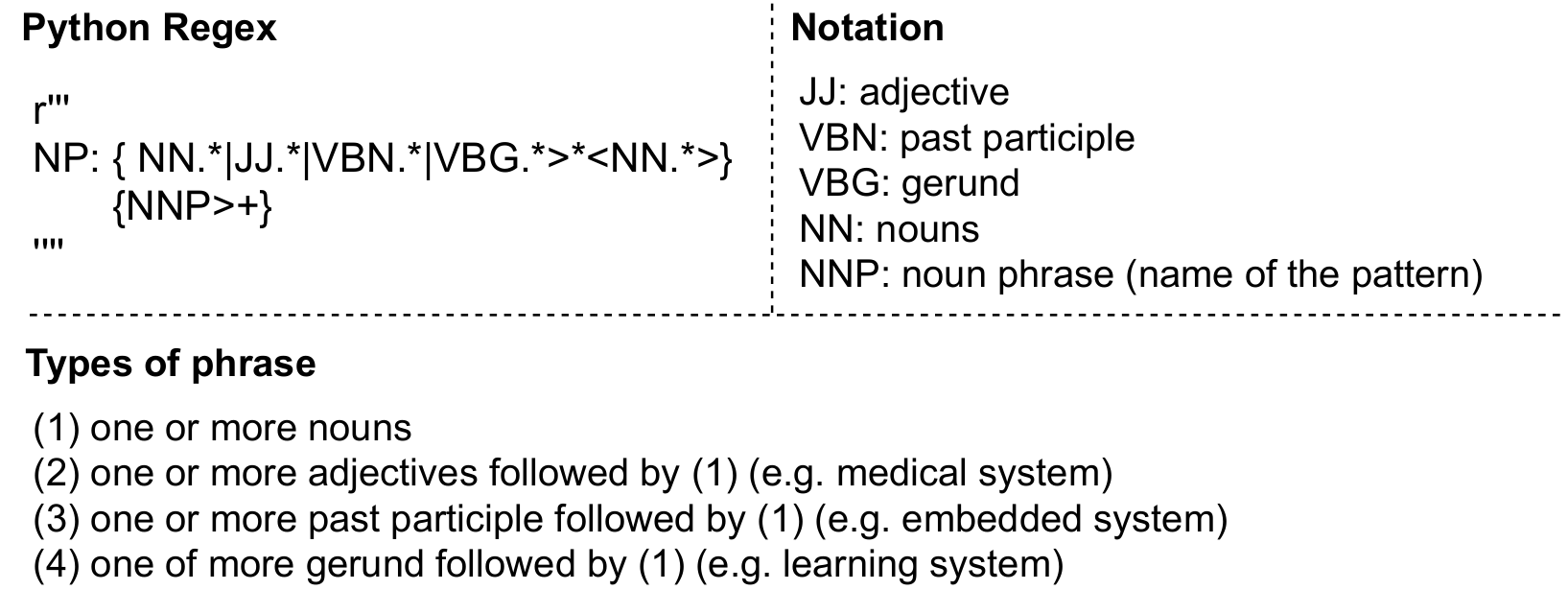}
    \caption{The Python regular expression of a linguistic pattern for extracting topics in a single document} 
    \label{fig:regex}
    \end{figure}
    Note that we use \texttt{nltk} for performing this process. The output of this step is the list of the tokens and the list of topics for each document $d \in \mc{D}$. The set of the all tokens is denoted as $\mc{W}$, and the set of the all topics is denoted as $\mc{T}$.

    \item \textbf{Step 2 - Estimate the weights of experts and documents given topics in $n$VSM}: The process includes four main steps with the corresponding functions in \textbf{generator.py}:
    \begin{enumerate}
        \item We use \texttt{generate\_tf()} to estimate the term frequencies (TFs) of $\mc{W}$ in each document $d \in \mc{D}$. For this estimation, we use \texttt{CountVectoerizer} in \texttt{scikit-learn}. The output of this function is the $|\mc{D}| \times |\mc{W}|$ Document-Token matrix (\textbf{DTM}) where each entry contains the TF of $w \in \mc{W}$ in $d$.
        
        \item We use  \texttt{generate\_dp\_matrix()} to estimate the weights of documents given $\mc{T}$ in $n$VSM \cite{kang2021expfinder}. The function estimates $n$TFIDF of each topic $t \in \mc{T}$ by integrating the $n$TF weighting and  the $n$IDF weighting. Intuitively, $n$TF estimates the frequency of $t$ by averaging TFs of tokens in $t$ where TF of each token is stored in \textbf{DTM}. In addition, $n$IDF \cite{nidf:2017} is the $N$-gram IDF weighting method that estimates the log-IDF, $\log\frac{|\mc{D}| \times df(t) + 1}{df(w_{1} \land w_{2} \land \ldots \land w_{n})^{2} + 1} +1$, of $t$ where $w_{1}, \ldots, w_{n}$ are $n$-constituent terms in $t$. The output of this step is the $|\mc{D}| \times |\mc{T}|$ Document-Phrase matrix (\textbf{DPM}) where each entry contains the $n$TFIDF weight of $t$ in $d$.
        
        \item Given $\mc{D}$, we use  \texttt{genenerate\_ed\_matrix()} to generate the $|\mc{X}| \times |\mc{D}|$ Expert-Document matrix (\textbf{EDM}) where each entry shows a binary relationship between $x \in \mc{X}$ and $d$ (e.g., 1 indicates that $x$ has the authorship on $d$, and 0 otherwise).

        \item We use  \texttt{generate\_pr\_matrix()} to estimate the weights of experts $\mc{X}$ and documents $\mc{D}$ given each topic $t \in \mc{T}$ in $n$VSM \cite{kang2021expfinder}. The weights of $\mc{X}$ are estimated by calculating matrix multiplication of $\textbf{EDM}^{|\mc{X}| \times |\mc{D}|}$ and $\textbf{DPM}^{|\mc{D}| \times |\mc{T}|}$ (e.g., \textbf{ETopM} \texttt{= numpy.matmul(}\textbf{EDM}, \textbf{DPM}\texttt{)} in Python). The output is the $|\mc{X}| \times |\mc{T}|$ Expert-Topic matrix (\textbf{ETopM}) where each entry contains the topic-sensitive weight of $x$ given $t$. The weights of $\mc{D}$ are represented by \textbf{DPM}.  Now, we denote \textbf{DPM} as the $|\mc{D}| \times |\mc{T}|$ Document-Topic matrix (\textbf{DTopM}) where each entry shows the topic-sensitive weight of $d$ given $t$. It is worth noting that \textbf{DPM} can be integrated with another factor (e.g., the average document frequencies of $\mc{T}$) to obtain different weights for \textbf{DTopM}. However, in our approach, we set $\textbf{DTopM} = \textbf{DPM}$.
        
    \end{enumerate}

    \item \textbf{Step 3 - Construct ECG}. We use  \texttt{generate\_ecg()} in \textbf{generator.py} to handle this step. The function receives $\mc{D}$ and builds an ECG using \texttt{DiGraph} in \texttt{networkx} to present a \textit{directed, weighted bipartite graph} that has expert nodes $V_{x}$ and document nodes $V_{d}$. The set of nodes in the graph is denoted as $V$ such that $V = V_x \cup V_d$. A directed edge points from a document node $v_{d} \in V_{d}$ to an expert node $v_{x} \in V_{x}$ if $x$ has published $d$. In this step, we also use \texttt{generate\_ed\_vector()} in \textbf{generator.py} to generate a $|V| \times 1$ Expert-Count vector ($\boldsymbol{c_{x}}$) and a $|V| \times 1$ Document-Count ($\boldsymbol{c_{d}}$) vector based on ECG. These vectors are used for the estimation of $\mu$CO-HITS in \textbf{Step 4}.

    \item \textbf{Step 4 - Reinforce expert weights using $\mu$CO-HITS}. We use \texttt{run\_expfinder()} in \textbf{trainer.py} to handle this step. The function receives \textbf{ETopM}, \textbf{DTopM}, ECG, $\boldsymbol{c_{x}}$ and $\boldsymbol{c_{d}}$, generated in \textbf{(Steps 2 and 3)} as parameters, and reinforces the estimation of expert weights given topics by integrating $n$VSM and $\mu$CO-HITS \cite{kang2021expfinder}. For each $t \in \mc{T}$, we perform the three steps:
    
    \begin{enumerate}
        \item \textbf{Generate the adjacency matrix of nodes and its transpose} - Given the ECG, we use  \texttt{to\_matrix()} in \texttt{networkx} to generate the $|V| \times |V|$ adjacency matrix of the graph $\textbf{M}$, and also construct its  transpose matrix $\textbf{M}^{\top}$. These matrices are required in the initialisation for running the $\mu$CO-HITS algorithm.
        
        \item \textbf{Normalise the weights of experts and documents given a topic} - We get topic-sensitive weights of $\mc{X}$ and $\mc{D}$ given $t$ from \textbf{ETopM} and \textbf{DTopM}, respectively. The output of this includes the $|\mc{X}| \times 1$ Expert-Topic ($\boldsymbol{\alpha_{x}}$) and $|\mc{D}| \times 1$ Document-Topic ($\boldsymbol{\alpha_{d}}$) vectors where each entry shows the topic-sensitive weight of an expert and a document given $t$, respectively. Then, we normalise these vectors using L2 normalisation to scale their squares sum to 1 as the initialisation for running the $\mu$CO-HITS algorithm \cite{kleinberg1999authoritative}.
        
        \item \textbf{Reinforce expert weights given a topic} - We integrate $n$VSM and $\mu$CO-HITS through $k$ iterations to reinforce expert weights given $t$. $\mu$CO-HITS is the extension of the CO-HITS algorithm \cite{Hongbo2009} which contains two main properties such as \textit{average authorities} $\boldsymbol{a}$ and \textit{average hubs} $\boldsymbol{h}$ which show importance of $\mc{X}$ and $\mc{D}$, respectively, based on the ECG. In addition, these properties can be defined as \cite{kang2021expfinder}:
        \begin{align}
            \boldsymbol{a}(\mc{X}; t)^{k} &= (1 - \lambda_{x}) \boldsymbol{a}(\mc{X}; t)^{k - 1} + \lambda_{x} \left( \frac{ \textbf{M}^{\top} \boldsymbol{\cdot} \boldsymbol{h}(\mc{D}; t)^{k - 1} }{ \boldsymbol{c_{d}} } \right) \\
            \boldsymbol{h}(\mc{D}; t)^{k} &= (1 - \lambda_{d}) \boldsymbol{h}(\mc{D}; t)^{k - 1} + \lambda_{d} \left( \frac{ \textbf{M} \boldsymbol{\cdot} \boldsymbol{a}(\mc{X}; t)^{k} }{ \boldsymbol{c_{x}} }  \right)
        \end{align}
        where
        \begin{itemize}
            \item $\boldsymbol{a}(\mc{X}; t)^{k}$ and $\boldsymbol{h}(\mc{D}; t)^{k}$ are  $|V| \times 1$ vectors which contain the reinforced expert weights and reinforced document weights, respectively, given $t$ at $k^{\text{th}}$ iteration. As the initial weights of these vectors, we use the topic-sensitive weights of experts and documents estimated in $n$VSM. Thus, $\boldsymbol{a}(\mc{X}; t)^{0} = \boldsymbol{\alpha_{x}}$ and $\boldsymbol{h}(\mc{D}; t)^{0} = \boldsymbol{\alpha_{d}}$. By doing so, we integrate $n$VSM with $\mu$CO-HITS. Note that $\boldsymbol{a}(\mc{X}; t)^{0}$ is a $|\mc{X}| \times 1 $ vector, and $\boldsymbol{h}(\mc{D}; t)^{0}$ is a $|\mc{D}| \times 1$ vector. However, for easily implementing the HITS algorithm, we have transformed the dimension of these vectors into $|V| \times 1$ vectors where  additional entries hold the value of 0.
            
            \item $\lambda_{x} \in [0, 1]$ and $\lambda_{d} \in [0, 1]$ are \textit{parameters} for expert and document, respectively. These are used to control the impact of topic-sensitive weights on $\boldsymbol{a}$ and $\boldsymbol{h}$, respectively. Assigning lower values indicates the higher impact of topic-sensitive weights on $\boldsymbol{a}$ and $\boldsymbol{h}$.
            
            \item $\left( \frac{ \textbf{M}^{\top} \boldsymbol{\cdot} \boldsymbol{h}(\mc{D}; t)^{k - 1} }{ \boldsymbol{c_{d}} } \right)$ is the calculation for the \textit{average authorities}. The numerator performs  matrix multiplication between the $|V| \times |V|$ adjacency matrix $\textbf{M}^{\top}$ and the $|V| \times 1$ $\boldsymbol{h}$. The denominator is a $|V| \times 1$ counted vector $\boldsymbol{c_{d}}$ generated in \textbf{Step 3}. To calculate this in Python, we simply apply \texttt{numpy.matmul(}$\textbf{M}^{\top}$, $\boldsymbol{h}(\mc{D}; t)^{k - 1}$\texttt{)/}$\boldsymbol{c_{d}}$.
            
            \item $\left( \frac{ \textbf{M} \boldsymbol{\cdot} \boldsymbol{a}(\mc{X}; t)^{k} }{ \boldsymbol{c_{x}} }  \right)$ is the calculation for the \textit{average hubs}. The numerator performs  matrix multiplication between the $|V| \times |V|$ adjacency matrix $\textbf{M}$ and the $|V| \times 1$ $\boldsymbol{a}$. The denominator is a $|V| \times 1$ counted vector $\boldsymbol{c_{x}}$ generated in \textbf{Step 3}. To calculate this in Python, we simply apply \texttt{numpy.matmul(}$\textbf{M}$, $\boldsymbol{a}(\mc{X}; t)^{k}$\texttt{)/}$\boldsymbol{c_{x}}$.
        \end{itemize}
        After computing $\boldsymbol{a}$ and $\boldsymbol{h}$ at $k^{\text{th}}$ iteration, we apply L2 normalisation to both $\boldsymbol{a}$ and $\boldsymbol{h}$. 
        We use the obtained $\boldsymbol{a}(\mc{X}; t)$ after the final iteration to construct the $|\mc{X}| \times |\mc{T}|$ Expert-Topic matrix (\textbf{RETopM}) where each entry contains the reinforced weight of $x$ given $t$.
        
    \end{enumerate}

\end{enumerate}

%% file: examples.tex
\section{Illustrative examples}\label{sec:example}

\begin{figure*}[!t]
    \centering
    \includegraphics[trim=0cm 0cm 0cm 0cm, clip, width=350pt]{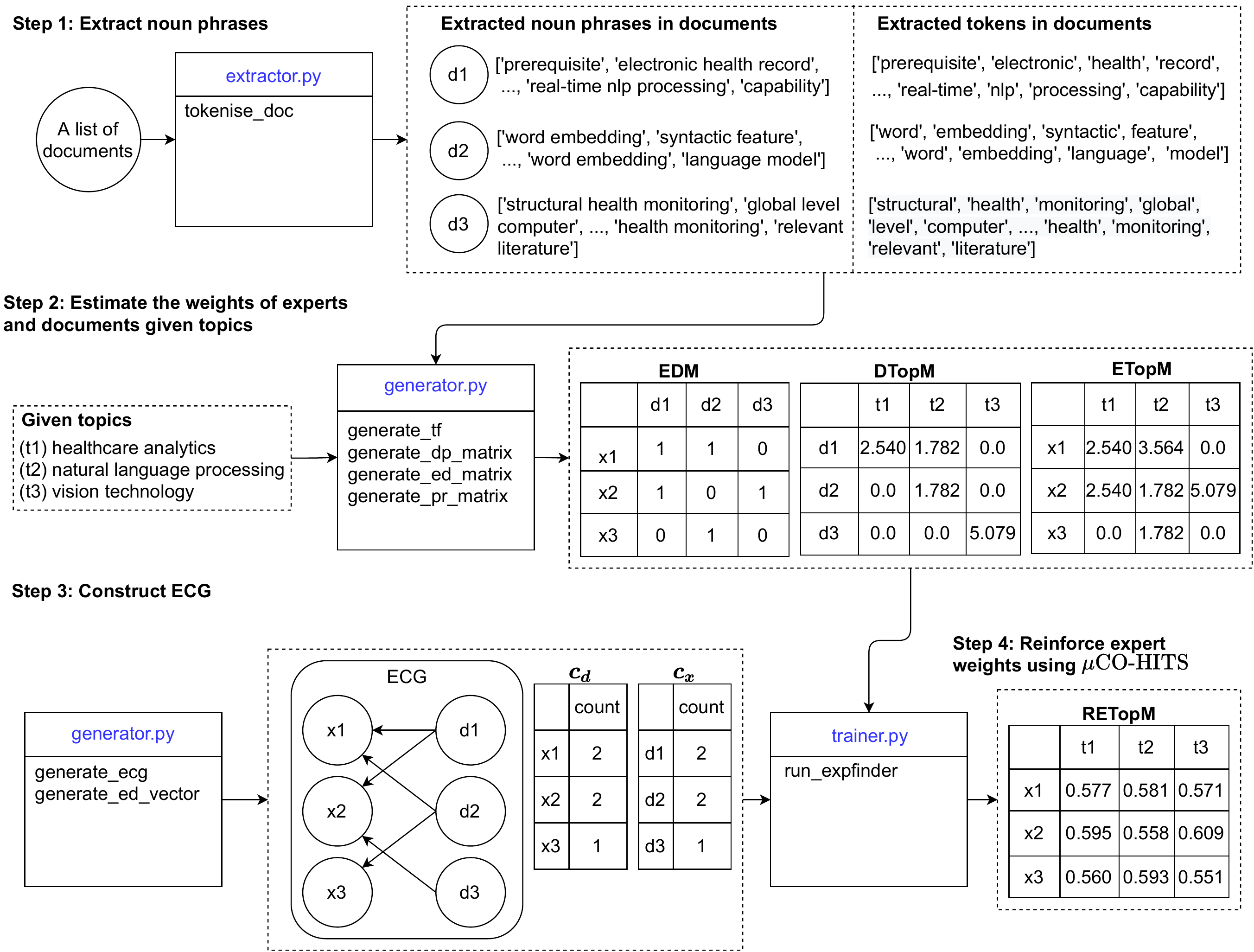}
    \caption{Illustrative examples for ExpFinder: the blue labels indicate module names of ExpFinder.} 
    \label{fig:process_example}
\end{figure*}

In this section, we illustrate how ExpFinder works. The input data\footnote{The example data are also provided in \href{https://github.com/Yongbinkang/ExpFinder/tree/main/data}{our Github repository}.} includes three experts (i.e., $x_{1}$, $x_{2}$ and $x_{3}$) and three documents (i.e., $d_{1}$, $d_{2}$ and $d_{3}$) as shown in Table \ref{tab:example}. Figure \ref{fig:process_example} presents the output examples of the steps in ExpFinder:

\begin{table*}[!h]
\scalebox{0.85}{
\setlength{\tabcolsep}{5pt}
\renewcommand{\arraystretch}{1}
    \begin{tabular}{p{0.12\linewidth} | p{0.1\linewidth} | p{0.85\linewidth} }
        \hline
        \textbf{Docs} & \textbf{Experts} & \textbf{Text}  \\
        \hline
        $d_{1}$ & $x_{1}$, $x_{2}$ & A \hl{\textbf{prerequisite}} for using \hl{\textbf{electronic} \textbf{health} \textbf{records}} (\hl{\textbf{EHR}}) \hl{\textbf{data}} within \hl{\textbf{learning} \textbf{health-care} \textbf{system}} is an \hl{\textbf{infrastructure}} that \textbf{enables} \hl{\textbf{access}} to \textbf{EHR} \hl{\textbf{data}} \textbf{longitudinally} for \hl{\textbf{health-care} \textbf{analytics}} and \hl{\textbf{real} \textbf{time}} for \hl{\textbf{knowledge} \textbf{delivery}}. Herein, we \textbf{share} our \hl{\textbf{institutional} \textbf{implementation}} of a \hl{\textbf{big} \textbf{data-empowered} \textbf{clinical}} \hl{\textbf{natural} \textbf{language} \textbf{processing}} (\textbf{NLP}) \hl{\textbf{infrastructure}}, which not only \textbf{enables} \hl{\textbf{healthcare} \textbf{analytics}} but also has \hl{\textbf{real-time} \textbf{NLP} \textbf{processing}} \hl{\textbf{capability}}. \\
        \hline
        $d_{2}$ & $x_{1}$, $x_{3}$ & \hl{\textbf{Word} \textbf{embedding}}, where \textbf{semantic} and \hl{\textbf{syntactic} \textbf{features}} are \textbf{captured} from \hl{\textbf{unlabeled} \textbf{text} \textbf{data}}, is a \hl{\textbf{basic} \textbf{procedure}} in \hl{\textbf{Natural} \textbf{Language} \textbf{Processing}} (\textbf{NLP}). In this \hl{\textbf{paper}}, we first \textbf{introduce} the \hl{\textbf{motivation}} and \hl{\textbf{background}} of \hl{\textbf{word} \textbf{embedding}} and its related \hl{\textbf{language} \textbf{models}}. \\
        \hline
        $d_{3}$ & $x_{2}$ & \hl{\textbf{Structural} \textbf{health} \textbf{monitoring}} at \textbf{local} and \hl{\textbf{global} \textbf{levels}} using \hl{\textbf{computer}} \hl{\textbf{vision} \textbf{technologies}} has \hl{\textbf{gained}} much \hl{\textbf{attention}} in the \hl{\textbf{structural} \textbf{health} \textbf{monitoring}} \hl{\textbf{community}} in research and \hl{\textbf{practice}}. Due to the \hl{\textbf{computer} \textbf{vision} \textbf{technology}} \hl{\textbf{application} \textbf{advantages}} such as \textbf{non-contact}, \hl{\textbf{long} \textbf{distance}}, \textbf{rapid}, \hl{\textbf{low} \textbf{cost}} and \hl{\textbf{labor}}, and \hl{\textbf{low} \textbf{interference}} to the \hl{\textbf{daily} \textbf{operation}} of \hl{\textbf{structures}}, it is \textbf{promising} to consider \hl{\textbf{computer} \textbf{vision} \textbf{structural}} \hl{\textbf{health} \textbf{monitoring}} as a \hl{\textbf{complement}} to the \hl{\textbf{conventional} \textbf{structural} \textbf{health}} \hl{\textbf{monitoring}}. This \hl{\textbf{article}} \textbf{presents} a \hl{\textbf{general} \textbf{overview}} of the \hl{\textbf{concepts}}, \hl{\textbf{approaches}}, and \hl{\textbf{real-life} \textbf{practice}} of \hl{\textbf{computer} \textbf{vision} \textbf{structural}} \hl{\textbf{health} \textbf{monitoring}} along with some \hl{\textbf{relevant} \textbf{literature}} that is rapidly \textbf{accumulating}. \\
        \hline
    \end{tabular}}
    \caption{The document dataset $\mc{D}$ used in the example: extracted phrases are highlighted in yellow, and extracted tokens are in bold.}
    \label{tab:example}
\end{table*}

\begin{enumerate}
    \item \textbf{Step 1 - Extract tokens and topics}: Given $\mc{D}$, we extract tokens $\mc{W}$ and topics $\mc{T}$. In this step, we set a maximum length of phrase to be 3 such that we only obtain phrases  that have less than or equal to 3 tokens. Additionally, we use the linguistic pattern presented in Section \ref{sec:func}. The output of this step contains the set of 50 unique topics $\mc{T}$ and the set of 85 unique tokens $\mc{W}$. For example, extracted topics in $d_{1}$ include some single-token topics (e.g., \texttt{prerequisite} and \texttt{capability}) and some multi-token topics (e.g., \texttt{real-time nlp processing} and \texttt{electronic health record}). 
    
    \item \textbf{Step 2 - Estimate the weights of experts and documents given topics} - Given $\mc{T}$ and $\mc{W}$, we generate three main matrices (i.e., $\textbf{EDM}$, $\textbf{DTopM}$ and $\textbf{ETopM}$) that will also be used in \textbf{Step 4}. To do this, we perform the following:
    
    \begin{itemize}
        \item Given $\mc{W}$, we generate $\textbf{DTM}^{3 \times 85}$ where each entry shows the TF of a token $w \in \mc{W}$ in a document $d \in \mc{D}$. For example, the $3 \times 1$ vector of  \texttt{healthcare}, $\textbf{DTM}_{*,\texttt{healthcare}}$,  is $(1, 0, 0)$ which shows it occurs only in $d_1$ (see also $\mc{D}$ in Table \ref{tab:example}). As another example, we obtain $\textbf{DTM}_{*,\texttt{analytics}} = (2,0,0)$ which denotes that \texttt{analytics} appears twice in $d_{1}$.
        
        \item Given $\mc{T}$ and $\textbf{DTM}^{3 \times 85}$, we generate $\textbf{DPM}^{3 \times 50}$ where each entry contains the weight of a phrase $t \in \mc{T}$ for a document calculated in $n$VSM. For example, suppose that  \texttt{health analytics} is denoted as $t_{1}$, we then calculate $n$TF of $t_1$ in $d_1$ as:
        \begin{align*}
            n\text{TF}(t_{1}, d_{1}) = \frac{\textbf{DTM}_{1,\texttt{healthcare}} +  \textbf{DTM}_{1,\texttt{analytics}}}{|t_{1}|} \
            =  \frac{(1 + 2)}{2} = 1.5
        \end{align*}
        where $|t_{1}|$ is a number of tokens in $t_{1}$. Then, we calculate the $N$-gram IDF of $t_{1}$ as:
        \begin{align*}
            n\text{IDF}(t_{1}) &= \log{\frac{|D| \cdot df(t_{1}) + 1}{df(\textbf{DTM}_{*,\texttt{healthcare}} \land \textbf{DTM}_{*,\texttt{analytics}})^{2} + 1}} + 1 \\
            &= \log{\frac{ 3 \times 1 + 1 }{ df((1, 0, 0) \land (2, 0, 0))^{2} + 1}} + 1 \\
            &= \log{\frac{4}{1^{2} + 1}} + 1 = 1.693.
        \end{align*}
        Here, $\land$ is implemented in  \texttt{NumPy}. Finally, we multiply $n\text{TF}(t_{1}, d_{1})$ with $n\text{IDF}(t_{1})$ to obtain $n$TFIDF of $d_{1}$ given $t_{1}$ as: $\textbf{DPM}_{1, t_{1}} = n\text{TFIDF}(t_{1}, d_{1}) = n\text{TF}(t_{1}, d_{1}) \times n\text{IDF}(t_{1}) = 1.5 \times 1.693 = 2.540$.
        
        \item Given $\mc{D}$, we generate $\textbf{EDM}^{3 \times 3}$ where each entry shows the authorship of an expert on a document. For example, $x_{1}$ is an author of $d_{1}$, and hence, the entry between $x_{1}$ and $d_{1}$ ($\textbf{EDM}_{1, 1}$) equals 1. Also, $\textbf{EDM}_{1, 3} = 0$ shows that $x_{1}$ is not an author of $d_{3}$ (See Table \ref{tab:example}).
        
        \item Given $\textbf{EDM}^{3 \times 3}$ and $\textbf{DPM}^{3 \times 50}$, we generate $\textbf{ETopM}^{3 \times 50}$ where each entry contains $n$TFIDF weight of an expert given a topic. As we explained in Section \ref{sec:func}, we assume that $\textbf{DTopM} = \textbf{DPM}$. Now, we demonstrate the calculation for the weights of experts $\mc{X}$ given $t_{1}$ ($\textbf{ETopM}_{*, t_{1}}$) in $n$VSM as:
        \begin{align*}
        \textbf{ETopM}_{*, t_{1}} &= \textbf{EDM}^{3 \times 3} \boldsymbol{\cdot} \textbf{DTopM}_{*, t_{1}} \\ 
        &= \begin{bmatrix} 1 & 1 & 0 \\ 1 & 0 & 1 \\ 0 & 1 & 0\end{bmatrix} \boldsymbol{\cdot} (2.540, 0, 0) =  (2.540, 2.540, 0)
        \end{align*}
    \end{itemize}
    Note that $\textbf{DTopM}^{3 \times 3}$ and $\textbf{ETopM}^{3 \times 3}$ are only used for the visualisation purpose. We use $\textbf{DTopM}^{3 \times 50}$ and $\textbf{ETopM}^{3 \times 50}$ for the estimation in \textbf{Step 4}.
    
    \item \textbf{Step 3 - Construct ECG}: Given $\mc{D}$, we generate an ECG which has three expert nodes and three document nodes, as shown in Figure \ref{fig:process_example}. The graph is also used to generate $3 \times 1$ vectors (i.e., $\boldsymbol{c_{x}}$ and $\boldsymbol{c_{d}}$) that are used for the estimation of $\mu$CO-HITS in \textbf{Step 4}. For example, $\boldsymbol{c_{d}}_{1} = 2$ indicates there are two documents (i.e., $d_{1}$ and $d_{2}$) pointing to $x_{1}$. Similarly, $\boldsymbol{c_{x}}_{3} = 1$ indicates that there is one expert (i.e., $x_{2}$) who has authorship on $d_{3}$.
    
    \item \textbf{Step 4 - Reinforce expert weights using $\mu$CO-HITS}: We use \texttt{run\_expfinder()} in \textbf{trainer.py} to reinforce expert weights given topics $\mc{T}$. The function receives $\textbf{DTopM}^{3 \times 50}$, $\textbf{ETopM}^{3 \times 50}$, ECG, $\boldsymbol{c_{x}}$ and $\boldsymbol{c_{d}}$, generated in \textbf{(Steps 2 and 3)} as parameters, and generate the $3 \times 50$ Expert-Topic matrix where each entry shows the reinforced weight of an expert given a topic. Now, we illustrate the estimation for the reinforced weight of $\mc{X}$ given $t_{1}$ as:
    \begin{itemize}
        \item Given 6 nodes in an ECG, we generate the adjacency matrix $\textbf{M}^{6 \times 6}$ and its transpose matrix $\textbf{M}^{\top}$ as:
        \begin{align*}
        \textbf{M} = \begin{bmatrix} 
            0 & 1 & 1 & 0 & 0 & 0 \\
            0 & 0 & 0 & 0 & 0 & 0 \\
            0 & 0 & 0 & 0 & 0 & 0 \\
            0 & 1 & 0 & 0 & 1 & 0 \\
            0 & 0 & 0 & 0 & 0 & 0 \\
            0 & 0 & 1 & 0 & 0 & 0
        \end{bmatrix}, 
        \textbf{M}^{\top} = \begin{bmatrix}
            0 &  0 &  0 &  0 &  0 &  0 \\
            1 &  0 &  0 &  1 &  0 &  0 \\
            1 &  0 &  0 &  0 &  0 &  1 \\
            0 &  0 &  0 &  0 &  0 &  0 \\
            0 &  0 &  0 &  1 &  0 &  0 \\
            0 &  0 &  0 &  0 &  0 &  0   
        \end{bmatrix}
        \end{align*}
        where rows and columns are labeled with the sequence $\boldsymbol{s}$ (i.e., $\boldsymbol{s} = (d_{1}, x_{1}, x_{2}, d_{2}, x_{3}, d_{3})$).
        
        \item We apply L2 normalisation for the $6 \times 1$ Expert-Topic ($\boldsymbol{\alpha_{x}}$) and the $6 \times 1$ Document-Topic ($\boldsymbol{\alpha_{d}}$) vectors. The output of each vector is as:
        \begin{align*}
            \boldsymbol{\alpha_{x}} &= \texttt{L2-normalize}(\textbf{ETopM}_{*, t_1}) = (0, 0.707, 0.707, 0, 0, 0) \\
            \boldsymbol{\alpha_{d}} &= \texttt{L2-normalize}(\textbf{DTopM}_{*, t_1}) = (1, 0, 0, 0, 0, 0)
        \end{align*}      
        
        \item We reinforce expert weights given $t_{1}$ in 5 iterations with $\lambda_{x} = 1$ and $\lambda_{d} = 0.7$. Here, we demonstrate the calculation of \textit{average authorities} $\boldsymbol{a}$ and \textit{average hubs} $\boldsymbol{h}$ at the first iteration ($k = 1$):
        \begin{align*}
            \boldsymbol{a}(\mc{X}; t_{1})^{1} &= (1 - \lambda_{x}) \boldsymbol{a}(\mc{X}; t_{1})^{0} + \lambda_{x} \left( \frac{ \textbf{M}^{\top} \boldsymbol{\cdot} \boldsymbol{h}(\mc{D}; t_{1})^{0} }{ \boldsymbol{c_{d}} } \right) \\
            &= 0 \cdot (0, 0.707, 0.707, 0, 0, 0) + 1.0 \cdot \left( \frac{(0, 2, 2, 0, 1, 0)}{(2, 1, 1, 2, 1, 1)} \right) \\
            &= (0, 2, 2, 0, 1, 0)
        \end{align*}
        \begin{align*}
            \boldsymbol{h}(\mc{D}; t_{1})^{1} &= (1 - \lambda_{d}) \boldsymbol{h}(\mc{D}; t_{1})^{0} + \lambda_{d} \left( \frac{ \textbf{M} \boldsymbol{\cdot} \boldsymbol{a}(\mc{X}; t_{1})^{1} }{ \boldsymbol{c_{x}} }  \right) \\
            &= 0.3 \cdot (1, 0, 0, 0, 0, 0) + 0.7 \cdot \left( \frac{(4, 0, 0, 3, 0, 2)}{(1, 2, 2, 1, 1, 1)} \right) \\
            &= (3.1, 0, 0, 2.1, 0, 1.4)
        \end{align*}
        where $\boldsymbol{a}(\mc{X}; t_{1})^{1}$ and $\boldsymbol{h}(\mc{D}; t_{1})^{1}$ are $6 \times 1$ vectors. At the end of the iteration, we normalise these vectors by applying the L2 normalisation technique as:
        \begin{align*}
            \boldsymbol{a}(\mc{X}; t_{1})^{1} &= \texttt{L2-normalize}(\boldsymbol{a}(\mc{X}; t_{1})^{1}) = (0, 0.667, 0.667, 0, 0.333, 0) \\
            \boldsymbol{h}(\mc{D}; t_{1})^{1} &= \texttt{L2-normalize} (\boldsymbol{h}(\mc{D}; t_{1})^{1}) =  (0.776, 0, 0, 0.525, 0, 0.35)
        \end{align*}
        After 5 iterations, we obtain $\boldsymbol{a}(\mc{X}; t_{1})^{5} = (0, 0.577, 0.595, 0, 0.56, 0)$ whose labels are presented by $\boldsymbol{s}$, and hence, we use $x_{1}$, $x_{2}$ and $x_{3}$ as indexes for obtaining a $3 \times 1$ vector (i.e., $\textbf{RETopM}_{*, t_{1}} = (0.577, 0.595, 0.56)$).
    \end{itemize}
    The output is $\textbf{RETopM}^{3 \times 50}$. If we use $t_{1}$ and the other two topics (i.e., \textbf{natural language processing} and \textbf{vision technology}, denoted as $t_{2}$ and $t_{3}$, respectively), we can generate $\textbf{RETopM}^{3 \times 3}$ in Figure \ref{fig:process_example}. This matrix can be used for two major tasks (1) finding the most expertise query for each expert (also known as expert profiling); and (2) finding the best expert for a given query (also known as expert finding). 

\end{enumerate}

%% file: impact.tex
\section{Impact and Conclusion} \label{sec:impact}
With the growth of expertise digital sources, expert finding is a crucial task that has significantly helped people to seek the services and guidance of an expert \cite{gonccalves2019automated}. ExpFinder is an ensemble model for expert finding that integrates $n$VSM with $\mu$CO-HITS to enhance the capability for expert finding over existing DLM, VSM and GM approaches. To our best knowledge, ExpFinder is the first attempt to provide the implementation of $n$VSM and $\mu$CO-HITS for expert finding.

The implementation of ExpFinder also provides functionalities that can be potentially useful for implementing other expert finding models. For example, our tokenisation module for extracting noun phrases using a linguistic pattern based on a part of speech (POS) can be easily customised based on researchers' purposes. The modules for building the presented Expert-Document matrix (\textbf{EDM}), Expert-Topic matrix (\textbf{ETopM}), and Document-Topic matrix (\textbf{DTopM}) can be usefully leveraged to represent relationships between experts and documents, experts and topics, and documents and topics. These relationships can be used to represent a collective information among experts, documents and topics and used to implement other graph-based expert finding models such as an author-document-topic (ADT) graph \cite{Gollapalli2013} and an expert-expert graph via topics.  

We highlight that ExpFinder is a state-of-the-art model, substantially outperforming the following widely known and latest models for expert finding: {document language models} \cite{Balog2009}, probabilistic-based expert finding model \cite{WISER2019}), {graph-based models} \cite{Hongbo2009, Gollapalli2013, Daniel2015}. Thus, the ones who want to extend ExpFinder can harness our implementation for further improvement of ExpFinder.

We presented the architecture and implementation detail of ExpFinder with an illustrative example. This would help researchers and practitioners to better understand how ExpFinder is designed and implemented with its core functionalities.

%% file: ExpFinder_SIMPA.bbl
\begin{thebibliography}{10}
\expandafter\ifx\csname url\endcsname\relax
  \def\url#1{\texttt{#1}}\fi
\expandafter\ifx\csname urlprefix\endcsname\relax\def\urlprefix{URL }\fi
\expandafter\ifx\csname href\endcsname\relax
  \def\href#1#2{#2} \def\path#1{#1}\fi

\bibitem{kang2021expfinder}
Y.-B. Kang, H.~Du, A.~R.~M. Forkan, P.~P. Jayaraman, A.~Aryani, T.~Sellis,
  Expfinder: An ensemble expert finding model integrating $n$-gram vector space
  model and $\mu$co-hits (2021).
\newblock \href {http://arxiv.org/abs/2101.06821} {\path{arXiv:2101.06821}}.

\bibitem{riahi2012finding}
F.~Riahi, Z.~Zolaktaf, M.~Shafiei, E.~Milios, Finding expert users in community
  question answering, in: Proceedings of the 21st International Conference on
  World Wide Web, 2012, pp. 791--798.

\bibitem{chuang2014combining}
C.~T. Chuang, K.~H. Yang, Y.~L. Lin, J.~H. Wang, Combining query terms
  extension and weight correlative for expert finding, in: 2014 IEEE/WIC/ACM
  International Joint Conferences on Web Intelligence (WI) and Intelligent
  Agent Technologies (IAT), Vol.~1, IEEE, 2014, pp. 323--326.

\bibitem{Balog2009}
K.~Balog, L.~Azzopardi, M.~de~Rijke, A language modeling framework for expert
  finding, Information Processing \& Management 45~(1) (2009) 1 -- 19.

\bibitem{Wang2015}
B.~Wang, X.~Chen, H.~Mamitsuka, S.~Zhu, Bmexpert\: Mining medline for finding
  experts in biomedical domains based on language model, IEEE/ACM Trans.
  Comput. Biol. Bioinformatics 12~(6) (2015) 1286--1294.

\bibitem{WISER2019}
P.~Cifariello, P.~Ferragina, M.~Ponza, {WISER: A semantic approach for expert
  finding in academia based on entity linking}, Information Systems 82 (2019) 1
  -- 16.

\bibitem{Hongbo2009}
H.~Deng, M.~R. Lyu, I.~King, {A Generalized CO-HITS Algorithm and Its
  Application to Bipartite Graphs}, in: Proceedings of the 15th ACM SIGKDD
  International Conference on Knowledge Discovery and Data Mining, 2009, p.
  239–248.

\bibitem{Gollapalli2013}
S.~D. Gollapalli, P.~Mitra, C.~L. Giles, Ranking experts using
  author-document-topic graphs, in: Proceedings of the 13th ACM/IEEE-CS Joint
  Conference on Digital Libraries, JCDL '13, 2013, pp. 87--96.

\bibitem{Daniel2015}
D.~Schall, A Social Network-Based Recommender Systems, Springer, 2015.

\bibitem{nidf:2017}
M.~Shirakawa, T.~Hara, S.~Nishio, {IDF for Word N-Grams}, ACM Trans. Inf. Syst.
  36~(1).

\bibitem{kleinberg1999authoritative}
J.~M. Kleinberg, Authoritative sources in a hyperlinked environment, Journal of
  the ACM (JACM) 46~(5) (1999) 604--632.

\bibitem{gonccalves2019automated}
R.~Gon{\c{c}}alves, C.~F. Dorneles, Automated expertise retrieval: A
  taxonomy-based survey and open issues, ACM Computing Surveys (CSUR) 52~(5)
  (2019) 1--30.

\end{thebibliography}
